\documentclass[10pt,a4paper]{article}
\usepackage{graphicx}
\usepackage{cite}

\newcommand{\ub}{\mathrm{ub}}
\renewcommand{\b}{\mathrm{b}}
\newcommand{\bin}{\mathrm{b,in}}
\newcommand{\ubin}{\mathrm{ub,in}}

\newcommand{\rhoB}{\rho_{\b}}
\newcommand{\rhoUb}{\rho_{\ub}}
\newcommand{\Dub}{D_{\ub}}
\newcommand{\vb}{v_{\b}}
\newcommand{\eps}{\epsilon_0}
\newcommand{\piad}{\pi_0}
\newcommand{\jB}{j_{\b}}
\newcommand{\JB}{\overline{J}_{\b}}

\newcommand{\opendiamond}{\mbox{$\diamondsuit$}}
\newcommand{\opencircle}{\mbox{\Large$\circ\,$}}  

\title{%
Molecular motor traffic in a half-open tube}
\author{Melanie J.I. M\"uller\thanks{Email: Melanie.Mueller@mpikg.mpg.de}, Stefan Klumpp and Reinhard Lipowsky\\%
\textit{\small Max Planck Institute for Colloids and Interfaces,14424 Potsdam, Germany}}
\date{}

\begin{document}

\maketitle

\begin{abstract}
The traffic of molecular motors which interact through mutual
exclusion is studied theoretically for half-open tube-like
compartments. These half-open tubes mimic the shapes of axons.
The mutual exclusion leads to traffic jams or density plateaus on the
filaments. A phase transition is obtained when the
motor velocity changes sign. We identify the relevant length
scales and characterize the jamming behavior using both analytical
approximations and Monte Carlo simulations of lattice models.
\end{abstract}


\section{Introduction}\label{secIntro}

Biological cells exhibit complex patterns of intracellular
traffic: vesicles shuttle between different cellular compartments,
others travel from the cell center to the periphery or vice versa,
but also filaments, RNA, chromosomes, and even viruses are permanently on the
move within cells \cite{AlbertsWalter2002,SchliwaWoehlke2003}. An
extreme example is the long-ranged transport of vesicles,
organelles, and proteins along the axons of nerve cells
\cite{GoldsteinYang2000}, which can
be up to a meter long. All this traffic
is based on the molecular motors kinesins, dyneins and
myosins which move along cytoskeletal filaments
\cite{SchliwaWoehlke2003,Howard2001,Schliwa2003}. These motors
catalyze a chemical reaction, the hydrolysis of ATP
(adenosinetriphosphate), and transform the free energy released
from this reaction into active movements and mechanical work. In
the following, we focus on processive cytoskeletal motors which
can make many chemo-mechanical steps while staying bound to a
filament, in particular on the large-scale traffic driven by these
motors \cite{LKN2001}.

These molecular motors exhibit movements on several length and
time scales which range between nanometers and millimeters or
centimeters (up to a meter in axons) and between microseconds and
days, respectively \cite{LipowskyKlumpp2005}: (I) The single step
of a motor typically has a size of the order of $10$~nm, and is
generated through the amplification of nanometer-sized
conformational changes in the catalytic domain of the motor. The
chemical cycle takes typically of the order of $10^{-2}$~s, but
the actual physical displacement is much faster, being almost
instantaneous on this time scale. (II) On scales of one or a few
microns, the motors perform active walks along cytoskeletal
filaments and move in a directed fashion with a velocity of $\sim$
$1~\mu$m/s. (III) On larger length and time scales which exceed a
few microns or a few seconds, respectively, the motors perform
peculiar 'motor walks', which consist of alternating sequences of
active directed walks along filaments and passive non-directed
diffusion in the surrounding fluid environment upon unbinding from
filaments \cite{Ajdari1995,LKN2001}. These motor walks are a
consequence of the fact that molecular motors function in a noisy
environment, and that the motor--filament binding energy can be
overcome by thermal fluctuations.

Theoretical investigations of molecular motors have largely
focused on movements on the intermediate scale (II), motivated by
single-molecule experiments developed during the last 15 years
which made the direct observation of the active motor walks
possible, see e.g.,
\cite{HowardVale1989,SvobodaBlock1993,ValeYanagida1996,VisscherBlock1999}.
These active walks have been studied using a variety of modeling
approaches such as various types of ratchet models (reviewed in
\cite{JuelicherProst1997,Lipowsky2000b,AstumianHaenggi2002}),
chemical kinetics models and discrete Brownian ratchets
\cite{Qian1997,FisherKolomeisky1999,Lipowsky2000a,LipowskyJaster2003}
and Brownian networks
\cite{LipowskyJaster2003,LipowskyKlumpp2005}. Only little has been
done on the dynamics of the single step, i.e., on movements in
regime (I), because the duration of a motor cycle still exceeds
the time scale accessible to Molecular Dynamics simulations.

The movements over large time and lengths scales, i.e.\ in regime
(III), have recently been studied in some detail. Our group
has introduced a class of lattice models which allows us to study
the combination of active walks along filaments and diffusion in a
rather generic way without making assumptions on the underlying
motor mechanisms, and, at the same time, to describe specific
motors by adapting the parameters to the measured transport
properties \cite{LKN2001}. These models also apply to various types
of 'motor particles' which consist of cargo particles with one or
several molecular motors
attached.

Using these models, many properties of
the large-scale motor walks can be calculated analytically
\cite{NieuwenhuizenKlumppLipowsky2002,NieuwenhuizenKlumppLipowsky2004}.
In addition, these lattice models can be used to study
motor--motor interactions which are easily incorporated.
The most obvious such interaction is the mutual exclusion
or hard-core repulsion which leads to traffic jams on crowded
filaments \cite{LKN2001}. In that case, these models constitute a
new class of driven diffusive systems or exclusion processes: In
contrast to driven lattice models studied in the context of
non-equilibrium phase transitions, e.g.
\cite{KatzSpohn1983,Krug1991,Schuetz2001}, and vehicular traffic
\cite{ChowdhurySchadschneider2000}, the driving, i.e.\ the active
movement is localized to the filaments and coupled to non-driven
diffusion through the binding/unbinding dynamics. These new driven
lattice models have been studied quite extensively both by our
group
\cite{LKN2001,KlumppLipowsky2003,KlumppLipowsky2004,KlumppNieuwenhuizenLipowsky2005}
and by several other groups
\cite{ParmeggianiFrey2003,PopkovSchuetz2003,EvansSanten2003,KleinJuelicher2005}.

One particularly interesting system is a tube-like cylindrical
compartment with filaments aligned in parallel to the cylinder axis in
an isopolar fashion, which mimics the geometry of an axon. Since
the lattice models for molecular motors are driven systems, the
boundary conditions are crucial for the resulting behavior.
In our previous work, we have studied tube-like compartments with
several types of boundary conditions: periodic boundary conditions
\cite{KlumppLipowsky2003}, closed boundaries
\cite{LKN2001,KlumppNieuwenhuizenLipowsky2005}, and open
boundaries, which are coupled to motor reservoirs
\cite{KlumppLipowsky2003}. The latter systems exhibit
boundary-induced phase transitions.

In this article, we consider a half-open tube which is coupled to
a motor reservoir at one end, but has a closed or reflecting wall at the
other end. We consider the two cases of motors moving towards the
open and towards the closed end. This geometry is inspired by the
geometry of the axon \cite{LKN2001}, which has a closed end at the axon terminal,
but, at the other end, is connected to the cell body, where the
motors are synthesized and which therefore acts as a motor
reservoir.

The paper is organized as follows: In section \ref{secModel}, we
introduce the model, in section \ref{secPhases}, we present the
general properties of these systems and discuss the phase
transition that is obtained when switching the direction of motor
motility. We then characterize the motor traffic jams in section
\ref{secJam} where we use an adsorption equilibrium approximation
to obtain analytical results. In the appendix, the same method is
applied to the closed tube system studied in
\cite{LKN2001,KlumppNieuwenhuizenLipowsky2005}. We close with a
short summary of our results.

\section{The model}\label{secModel}

\begin{figure}
\includegraphics[width=13cm]{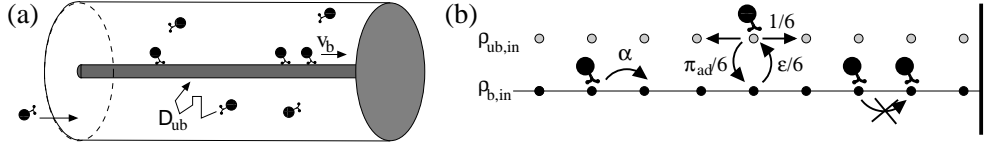}
\caption{%
Motion of motor particles along a filament within a half-open tube
(a) and the corresponding lattice model (b). At the left end, the
system is connected to a motor reservoir with fixed bound and
unbound densities $\rho_{\bin}$ and $\rho_{\ubin}$, while the
right end is closed.
Motor particles bound to the filament perform active movements with
velocity $\vb$, while unbound motors perform symmetric diffusion with
diffusion constant $\Dub$. In the lattice models, these movements are
described by the forward step probability $\alpha$ on the filament and
the symmetric hopping probability $1/6$ to each neighbour site for
unbound motors. In addition, unbound motors which reach the
filament bind to it with probability $\pi_{\mathrm{ad}}$, and bound motors
unbind from the filament through steps to adjacent non-filament sites which
occur with probability $\epsilon/6$ per non-filament neighbour site. The
motors interact via hard core exclusion.%
}
\label{FigLatticeModel}
\end{figure}

To mimic transport in an axon, we study the traffic of
motor particles in a half-open cylindrical tube with a single filament
located along its cylinder axis. The tube has length $L$ and
radius $R$. The bound motors move along the filament and the
unbound motors diffuse freely within the cylinder, see figure
\ref{FigLatticeModel}. At the right end, the tube is closed, i.e.\
no motors can leave or enter the system. This mimics the synaptic
terminal of the axon. At the left end, the system is coupled to a
motor reservoir which represents the cell body and provides fixed
bound and unbound motor densities $\rho_{\bin}$ and
$\rho_{\ubin}$, respectively.

Each motor can be in two states: bound to the filament, where it
moves actively to the right with velocity $\vb$, and unbound,
where it performs symmetric diffusion with diffusion constant
$\Dub$. The filament is taken to lie along the $x$-axis, so that the
system is characterized by the bound motor density $\rhoB(x)$ and
the unbound motor density $\rhoUb(x)$. As in
\cite{KlumppLipowsky2003,KlumppNieuwenhuizenLipowsky2005},
we neglect the variation
of the unbound motor density $\rhoUb$ with the transverse
coordinates $y$ and $z$.

The motors can unbind from or bind to the filament. Since the
motors are strongly attracted by the filament, the binding rate
$\piad$ is taken to be large compared to the unbinding rate
$\eps$. When motors come close to each other, they may interact.
In our simple model we include only hard core exclusion which
prevents motors from occupying the same site. In a mean field
treatment, this can be taken into account by using exclusion
factors of the form $(1-\rhoB)$ or $(1-\rhoUb)$.

We are interested in the stationary states of the system. Since the
right tube end is closed, the net current vanishes in the
stationary state. Thus, the directed bound current of motors, $\vb
\rhoB(1-\rhoB)$, and the unbound diffusive motor current, $-\phi
\Dub \frac{\partial}{\partial x}\rhoUb$, must balance to give zero
total current:
\begin{equation}\label{netCurrent}
    \vb \rhoB (1-\rhoB) - \phi \Dub \frac{\partial}{\partial x} \rhoUb = 0.
\end{equation}
Here the unbound diffusive current has been integrated over the
tube cross section. The prefactor $\phi$ describes the area available
for unbound diffusion. For large radii R, $\phi\approx\pi R^2$.
Furthermore, in the stationary state,
the in- and outgoing currents balance on any filament site which
leads to
\begin{equation}\label{Kirchhoff}
    \frac{\partial}{\partial x}\left[ \vb \rhoB(1-\rhoB)\right]
        = \piad\rhoUb(1-\rhoB) - \eps\rhoB(1-\rhoUb)
\end{equation}
where the small diffusive part of the bound current has been neglected.
The change of the bound current drives the system out of
adsorption equilibrium for which the binding term
$\piad\rhoUb(1-\rhoB)$ and the unbinding term
$\eps\rhoB(1-\rhoUb)$ would be equal, i.e.\
\begin{equation}\label{adsorpEquil}
     \piad\rhoUb(1-\rhoB) = \eps\rhoB(1-\rhoUb).
\end{equation}
For simplicity, we assume that the boundary densities at the left
boundary, $\rho_{\mathrm{b,in}}$ and $\rho_{\mathrm{ub,in}}$,
fulfill this adsorption equilibrium condition (\ref{adsorpEquil}).

In addition to studying equations
(\ref{netCurrent}) and (\ref{Kirchhoff}) by analytical approximations,
we use Monte Carlo simulations of lattice models as in
\cite{LKN2001,KlumppLipowsky2003} to obtain the stationary density
and current profiles. In the simulations, the motors are
represented by random walkers on a cubic lattice with lattice constant
$\ell$, see figure \ref{FigLatticeModel}(b). A line of lattice sites represents the
filament. Motors at the filament sites perform a biased random
walk, while motors at non-filament sites perform symmetric random
walks. The jump probabilities per unit time $\tau$ are $\alpha$, $\beta$ and
$\gamma$ for a forward, a backward and no step at filament sites,
respectively, and $1/6$ for steps to each neighbour site at non-filament
sites. In the following, we use $\beta=0$ for simplicity. Unbinding from
the filament occurs with probability $\epsilon/6$ to each adjacent
non-filament site, and unbound motors which reach the filament bind to
it with probability $\pi_{\mathrm{ad}}$.
Exclusion is implemented by rejecting all hopping attempts
to sites which are already occupied by another motor.
The hopping rates of the lattice models are matched to the transport
properties of the motors via $\tau=\ell^2/(6\Dub)$, $\vb=\alpha \ell$,
$\piad = 4\pi_{\mathrm{ad}}/6$ and $\eps=4\epsilon/6$, see \cite{LKN2001,KlumppLipowsky2003} and the appendix C in \cite{LipowskyKlumpp2005}.

\section{Steady states and relevant length scales}\label{secPhases}

\begin{figure}
\includegraphics[width=13cm]{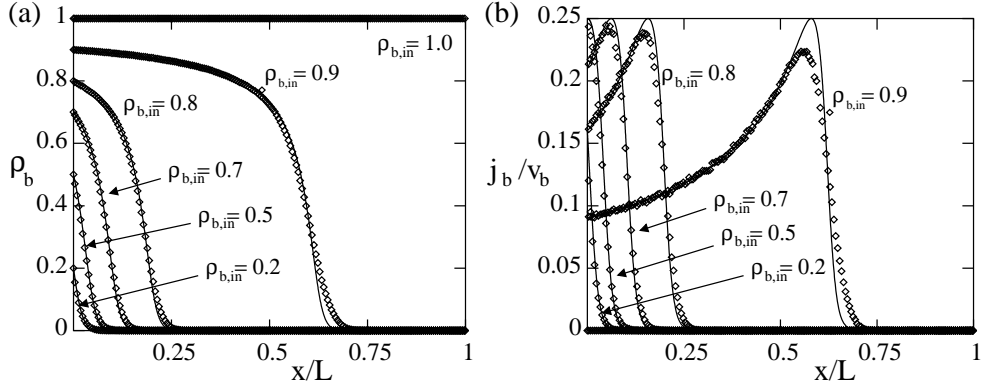}
\caption{%
(a) Bound density profiles $\rhoB$ and (b) current densities $\jB$
for different boundary densities $\rho_{\bin}$ as obtained from
Monte Carlo simulation ($\opendiamond$) and from the adsorption
equilibrium approximation (solid lines). The
parameters are $\vb=-0.001$, $\eps=2/300$, $\piad=2/3$,
$\phi=4$.%
} \label{figProfilesCurrents}
\end{figure}

When the motors on the filament move to the right, i.e.\
for $\vb>0$, they jam up at the closed right end until the tube far
from the left boundary is completely filled with motors. In the
bulk of a long tube, one thus has bound and unbound motor
densities $\rhoB = 1$ and $\rhoUb = 1$. Indeed, these density
values are a fixed point of equations (\ref{netCurrent}) and
(\ref{Kirchhoff}). Within a boundary region at the left end, the
densities cross over to the boundary values $\rho_{\mathrm{b,in}}$
and $\rho_{\mathrm{ub,in}}$.

Likewise, if the motors on the filament move to the left, i.e.\
for $\vb<0$, they are driven out of the tube at the left end, so
that the bulk is left empty with $\rhoB=0$ and $\rhoUb=0$, which
is another fixed point of equations (\ref{netCurrent}) and (\ref{Kirchhoff}).
However, a boundary region at the left end survives, fed by ingoing
diffusion. Typical motor density profiles for this case are shown in figure
\ref{figProfilesCurrents}(a). They display a "jam" region at the left
end, separated from the "empty" bulk region by a localized domain wall or shock.
The system is thus characterized by two length scales:
the bulk length scale $\xi$ on which the densities approach their the bulk
values and which is independent of the boundary densities, and the
jam length $L_{*}$ which describes the width of the jammed
boundary region and which, of course, strongly depends on the
boundary densities.

On tuning the motor velocity $\vb$, the system displays two
phases, a high and a low density phase, which are dominated by the
closed right end. Taking the bulk density as order parameter, one
has a phase transition at $\vb=0$. As will be shown both in analytical approximation and in Monte Carlo simulation, both characteristic length scales $\xi$ and $L_{*}$ diverge with
the power law $1/\vb$ as the motor velocity $\vb$ approaches zero.

In an experiment, this limit can easily be realized by reducing the
concentration [ATP] of the motor fuel ATP, since for small [ATP],
$\vb\sim$ [ATP]. Changing the sign of $\vb$ is much more difficult, one possibility is to
use motor particles driven by motors of two species with opposite
directionality, so that one could invert the movement by activating one
motor species while deactivating the other. If these motor particles switch
stochastically between the two directions, one could change the sign of
their average velocity by influencing the switching rates through regulatory
molecules. For example, tau proteins which strongly suppress movements towards
the synapse in axons can induce the retreat of vesicles into the cell
body \cite{MandelkowMandelkow2004}.

\begin{figure}
\includegraphics[width=13cm]{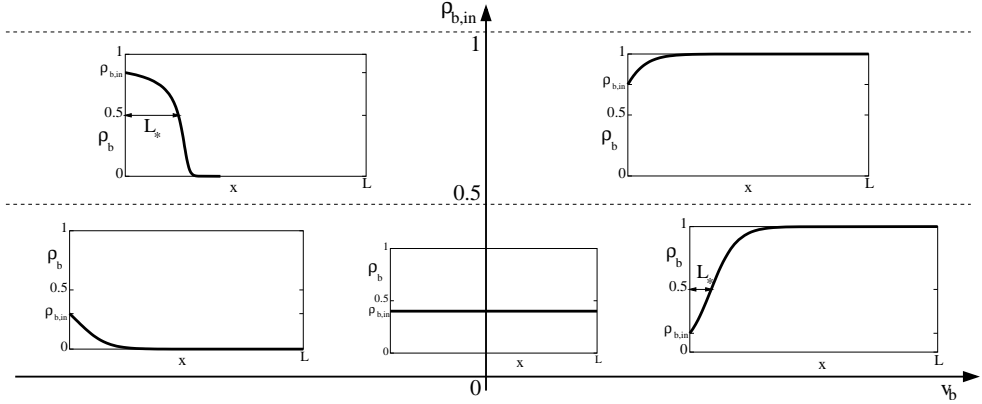}
\caption{%
Phase diagram for the half-open tube. For motors moving to the left, i.e.\ $\vb<0$, the system is essentially empty with bulk density $\rhoB=0$, while for motors moving to the right, i.e.\ $\vb>0$, it is essentially full with bulk density $\rhoB=1$. At the phase transition $\vb=0$, the system is dominated by the left boundary density and has the bulk density $\rhoB=\rho_{\bin}$. For $\vb<0$, the motors form a traffic jam with length $L_*$ at the left end, provided that the boundary density $\rho_{\bin}$ is larger than $1/2$. Likewise, for $\vb>0$ and $\rho_{\bin}<1/2$, a 'jam of holes', i.e.\ a region with low density, separates the (completely filled) bulk of the system from the left boundary.%
}\label{FigPhaseDiag}
\end{figure}

For $\vb=0$, i.e.\ when there is no active motion of the motors,
the system is in thermodynamic equilibrium, and the adsorption
equilibrium as described by equation (\ref{adsorpEquil}) is valid for
all $x$ along the filament. Thus the bulk densities are
$\rhoB=\rho_{\bin}$ and $\rhoUb=\rho_{\ubin}$. This case is the
only one where the bulk values are dominated by the open left
filament end. A phase diagram is shown in figure \ref{FigPhaseDiag},
the control parameters are the motor velocity $\vb$ and the left boundary
density $\rho_{\bin}$.

To examine the bulk behaviour, equations
(\ref{netCurrent}) and (\ref{Kirchhoff}) are linearized around the
appropriate fixed point. In the case of motors moving to the left
($\vb<0$) with bulk values $\rhoB = \rhoUb = 0$, the
linearization leads to an exponential density profile with
\begin{equation}\label{bulkApprox}
    \rhoB \approx \mathcal{N}\exp(-x/\xi) \quad \mathrm{and} \quad
    \rhoUb \approx \left(\frac{\eps}{\piad} -
    \frac{\vb}{\piad\xi}\right)\rhoB
\end{equation}
with the bulk length scale
\begin{equation}\label{xi}
\xi = 2 \Delta x_{\b}\left[ 1 - \sqrt{1 + 4\left(\frac{\Delta
x_{\b}}{\Delta x_{\ub}}\right)^2}\right]^{-1}
    \approx - \frac{(\Delta x_{\ub})^2}{\Delta x_{\b}}
     = - \frac{\eps}{\piad}\frac{\phi\Dub}{\vb},
\end{equation}
where
\begin{equation}
    \Delta x_{\b} = \frac{\vb}{\eps}\quad \mathrm{and}\quad
    \Delta x_{\ub} = \sqrt{\frac{\phi \Dub}{\piad}}
\end{equation}
are the average walking distance on the filament before unbinding
and the average diffusion distance before binding of a motor,
respectively. The approximation in equation (\ref{xi}) is valid
for small motor velocities $\vb$. The exponential behaviour
near the bulk value is in agreement with simulations.

In the case of motors moving to the right, the same reasoning
leads to an exponential approach of the bulk values $\rhoB=1$ and
$\rhoUb=1$ with the length scale
\begin{equation}\label{xiHD}
\xi = - 2\Delta x_{\b,\mathrm{h}}
    \left[ 1 - \sqrt{1 + 4\left(\frac{\Delta x_{\b,\mathrm{h}}}{\Delta x_{\ub,\mathrm{h}}}\right)^2}  \right]^{-1}
    \approx \frac{(\Delta x_{\ub,\mathrm{h}})^2}{\Delta x_{\b,\mathrm{h}}}
     = \frac{\piad}{\eps}\frac{\phi\Dub}{\vb}
\end{equation}
where the last expressions are again valid for small motor
velocity $\vb$, and where $\Delta x_{\b,\mathrm{h}} =
\vb/\piad$  and $\Delta x_{\ub,\mathrm{h}} =
\sqrt{\phi \Dub/\eps}$ are the average
walking distance on the filament before unbinding and the average
diffusion distance before binding of a hole, respectively.  This
result can be obtained directly from (\ref{xi}) by exploring
particle-hole symmetry, i.e.\ by substituting the left by the
right boundary condition, $\rhoB$ by $1-\rhoB$, $\rhoUb$ by
$1-\rhoUb$, $\vb$ by $-\vb$ and $x$ by $L-x$, and by exchanging $\piad$ and $\eps$.

According to equations (\ref{xi}) and (\ref{xiHD}) the bulk length
scale $\xi$ diverges in the limit of small $\vb$ with the
power law $1/\vb$. The same power law is obtained both if $\vb=0$ is
approached from above and from below, but with different amplitudes.


\section{The jam region}\label{secJam}

In the following, we will focus on motors moving to the left,
i.e.\ on the case $\vb<0$. The case of motors moving to the right
can easily be deduced as above by invoking particle-hole symmetry.
Left-moving motors jam up in front of the left boundary, leading
to a boundary dominated length scale in the system, the jam length
$L_{*}$. In the jam region, the motors move slowly and thus "have
time" to equilibrate with the unbound motors before moving on.
This separation of time scales leads to approximate adsorption
equilibrium
\begin{equation}\label{adsorpEquil2}
     \piad\rhoUb(1-\rhoB) \approx \eps\rhoB(1-\rhoUb)
\end{equation}
at every filament site $x$.

\subsection{Density profile and jam length}\label{secJamDens}

With the adsorption equilibrium approximation as given by equation
(\ref{adsorpEquil2}), one can eliminate
$\rhoUb$ from the current balance equation (\ref{netCurrent}) which leads to
\begin{equation}\label{rhoBPrime}
    \frac{\partial}{\partial x}\rhoB = \frac{V}{L} \frac{(1-K)^2}{K} \rhoB (1-\rhoB)\left(\frac{1}{1-K}-\rhoB\right)^2
\end{equation}
where the dimensionless constants
\begin{equation}
K=\frac{\eps}{\piad} \quad \mathrm{and} \quad V=\frac{L\vb}{\phi\Dub}
\end{equation}
have been introduced. The desorption constant $K$ is the ratio of
unbinding to binding rate, while the dynamic constant $V$ is the
ratio of the time $t_{\ub} = L^2/(\phi\Dub)$ needed to diffuse
over the filament length $L$ to the time $t_{\b} = L/\vb$ needed
to walk this distance.  Separation of variables, decomposition into partial fraction
and subsequent integration of equation (\ref{rhoBPrime}), using
the boundary condition $\rhoB(0)=\rho_{\bin}$, leads to an
implicit equation for the bound density $\rhoB$:
\begin{equation}\label{implicitRho}
      \frac{x}{L} = \frac{1}{V}\left[ g(\rhoB(x),K)  -  g(\rho_{\bin},K)\right].
\end{equation}
The function
\begin{equation}\label{gfunc}
g(\rho,K) =  \frac{-1}{\frac{1}{1-K}-\rho} + K \ln(\rho) -
\frac{1}{K}\ln(1-\rho)
            + \frac{1-K^2}{K}\ln\left|\frac{1}{1-K}-\rho\right|
\end{equation}
is a continuous, monotonically increasing, and thus invertible
function for $0\leq \rho\leq 1$ and respects particle-hole
symmetry which is equivalent to $g(1-\rho,1/K) = - g(\rho,K)$. In the jam region, the
profiles obtained from equation (\ref{implicitRho}) agree nicely
with the Monte Carlo profiles, while they differ in the shock and
the bulk region, where adsorption equilibrium is no longer a good
approximation, see figure \ref{figProfilesCurrents}(a). In figure
\ref{figProfilesCurrents}(b), we also show the corresponding
profiles of the bound current $\jB=\vb\rhoB(1-\rhoB)$ which
exhibits a maximum close to the domain wall between the jammed and
the bulk region.

\begin{figure}
\includegraphics[width=13cm]{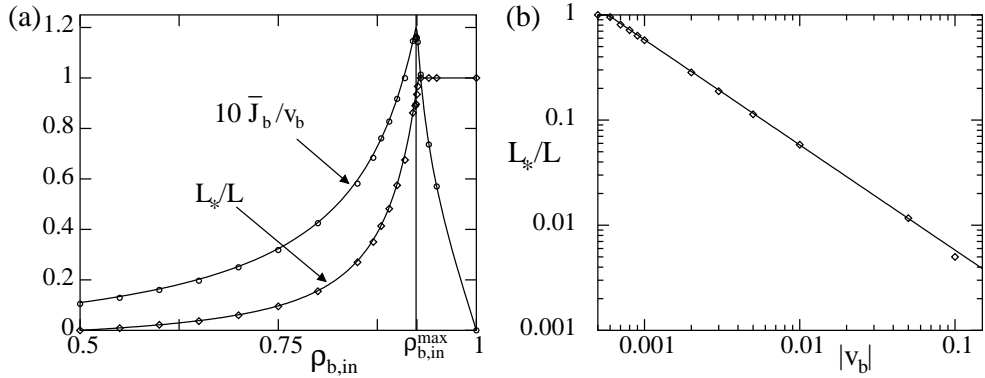}
\caption{%
Jam length $L_{*}$ as a function (a) of the boundary density
$\rho_{\bin}$ (with fixed velocity $\vb=-0.001$) and (b) of the motor
velocity $\vb$ (with fixed $\rho_{\bin}=0.9$). Both
Monte Carlo ($\opendiamond$) and analytic
results (solid lines) from the jam region approximation (\ref{xS}) are shown.
The other parameters are as in figure \ref{figProfilesCurrents}. In
(a), we also display the averaged bound current $\JB$ as obtained
from simulations ($\opencircle\hspace{-1ex}$) and from adsorption
equilibrium approximation (\ref{Jbtot0}) with its maximum at
$\rho_{\bin}^{\max}$. %
} \label{figJamLength}
\end{figure}

We therefore use the position of the current maximum as a
definition for the jam length $L_*$. In the jam region the
current is low due to a too high motor density while on the other
side of the shock there are too few motors, also leading to a
small current. As the maximal current is obtained for $\rhoB=1/2$,
the jam length $L_*$ is given by
\begin{equation}\label{xS}
 \frac{L_*}{L} = \frac{1}{V} \left[ g(\frac{1}{2},K) - g(\rho_{\bin},K) \right].
\end{equation}
Note that the jam length $L_*$ is positive only if
$\rho_{\bin}\geq 1/2$. For smaller boundary densities $L_*<0$
which implies that the domain wall is located outside the tube
which then does not exhibit a motor traffic jam. The jam lengths
determined from equation (\ref{xS}) are shown in figure
\ref{figJamLength} and agree well with the corresponding
simulation results. As expected, the jam length increases with the
boundary density $\rho_{\bin}$. It diverges logarithmically,
$L_*\approx L/(VK) \ln(1-\rho_{\bin})$, as $\rho_{\bin}$ approaches $1$ (of
course this divergence is truncated by the finite system size). If
the velocity $\vb$ approaches zero, the jam length diverges as
$1/\vb$ and the jammed region spreads over the whole system in
agreement with the fact that for $\vb=0$ the left boundary
dominates the whole system, as has been discussed in section
\ref{secPhases} above. The exponent $-1$ of the divergence for $\vb\rightarrow 0$, which has been obtained using mean field and adsorption equilibrium approximations, is confirmed by the Monte Carlo simulation results, see figure \ref{figJamLength}(b).

\subsection{The case $K=1$}

In the special case $K=1$, where binding to the filament and
unbinding from the filament occur with the same rate, the implicit
equation (\ref{implicitRho}) for the density profile can be
inverted and leads to the density profile
\begin{equation}\label{ProfileK1}
    \rhoB(x) = \left[ 1 + \frac{1-\rho_{\bin}}{\rho_{\bin}}\;\exp\left({-V \frac{x}{L}}\right)\right]^{-1}.
\end{equation}
The jam length $L_*$ in this case is given by
\begin{equation}
    \frac{L_*}{L} = \frac{1}{V}\ln\left(\frac{1-\rho_{\bin}}{\rho_{\bin}}\right).
\end{equation}

\subsection{The average bound current}

In order to characterize the overall transport in the system we
consider the average bound current \cite{LKN2001} which is defined
by
\begin{equation}
\JB = \frac{1}{L}\int_{0}^{L} \jB(\rhoB(x))\, dx.
\end{equation}
Since the bound density is essentially zero in the bulk, the bound
current $\jB(\rhoB)=\vb\rhoB(1-\rhoB)$ has large absolute values
only in the boundary region near the left end. It is thus
appropriate to use the jam region approximation of section
\ref{secJamDens} to calculate the average bound current $\JB$,
which leads to
\begin{eqnarray}\label{Jbtot0}
    \frac{\JB}{\vb}
    = \frac{1}{V} \frac{K}{(1-K)^2}
    \left[ \frac{1}{\frac{1}{1-K}-\rhoB(L)} - \frac{1}{\frac{1}{1-K}-\rho_{\bin}}\right].
\end{eqnarray}
For large system size $L$ the right boundary density
vanishes, and thus the average bound current is
given by
\begin{equation}
    \frac{\JB}{\vb} \approx \frac{1}{V} \frac{K}{(1-K)}\left[ 1 - \frac{1}{1-(1-K)\rho_{\bin}}\right].
\end{equation}
Interestingly, as $V\sim\vb$, the current $\JB$ in the
thermodynamic limit does not depend on the motor velocity $\vb$.
This is because the jam length $L_{*}$ decreases as $1/\vb$, thus
cancelling the expected $\JB\sim\vb$ behaviour.

The average bound current from equation (\ref{Jbtot0}) agrees well
with the current from the Monte Carlo simulation, see figure
\ref{figJamLength}(a). As a function of the left boundary density
$\rho_{\bin}$, it displays a maximum absolute value at a density
$\rho_{\bin}^{\max}$. This density thus optimizes the motor
transport in the system. It can be calculated via
\begin{eqnarray}\label{currentDerivative}
  0 &=& \frac{\partial \JB}{\partial \rho_{\bin}}
    = \frac{1}{V}\left[ \jB\left(\rhoB(L)\right)g'(\rhoB(L))\frac{\partial\rhoB(L)}{\partial\rho_{\bin}}
        -\jB\left(\rho_{\bin}\right)g'(\rho_{\bin})\right] \nonumber\\
    &=&\frac{1}{V}\left[ \jB\left(\rhoB(L)\right)-\jB\left(\rho_{\bin}\right)\right]\,
    g'(\rho_{\bin}).
\end{eqnarray}
The last expression follows from
\begin{equation}
    0 = g'(\rhoB(L))\frac{\partial\rhoB(L)}{\partial\rho_{\bin}} - g'(\rho_{\bin})
\end{equation}
which is obtained by differentiating (\ref{implicitRho}), taken at
$x=L$, with respect to $\rho_{\bin}$. As $g'(\rho_{\bin})>0$ for
$0\leq\rho_{\bin}\leq 1$, a current extremum can occur only for
\begin{equation}\label{currentMaxCondition1}
    \jB\left(\rho_{\bin}\right) = \jB\left(\rhoB(L)\right).
\end{equation}
Since equality of the densities at the left and right end occurs only
for a completely full or completely empty tube, the condition for
extremal current is
\begin{equation}\label{CurrentMaxCondition2}
    \rho_{\bin} = 1-\rhoB(L).
\end{equation}
Using this condition in equation (\ref{implicitRho}) with $x=L$,
one obtains the density that extremizes the average current from
\begin{equation}\label{CurrentMaxCondition3}
    V = g(\rho_{\bin}^{\max}) - g(1-\rho_{\bin}^{\max})
\end{equation}
The theoretical calculation agrees well with the simulation
results, see figure \ref{figJamLength}. The current extremum
occurs when the tube is approximately full, i.e.\ when the jam
length $L_{*}$ is only slightly smaller than the tube length $L$.
This is consistent with the assumption that the average current is
mainly supported by the jam region, on which the use of the
adsorption equilibrium approximation for the calculation of the
average current is based.

In the limit of large $L$, one must thus jam
up an infinite tube, for which one needs a boundary density
$\rho_{\bin}=1$. For this case, equation
(\ref{CurrentMaxCondition3}) can be inverted approximately which
leads to
\begin{equation}
    \rho_{\bin}^{max} \approx 1 - K^{\frac{1-K^2}{1+K^2}} \mathrm{e}^{\frac{-(1-K)^2}{1+K^2}}
     \exp{\left[-\frac{K}{1+K^2}\frac{|\vb|}{\Dub\phi}L\right]},
\end{equation}
i.e.\ the maximum density $1$ is approached exponentially for large $L$.


\section{Summary and conclusions}\label{secSummary}

We have studied the molecular motor traffic in half-open tube-like
compartments and characterized the traffic jams which result from
the mutual exclusion of motors. In particular, we have obtained
analytical results for the jam length and for the parameters which
optimize the traffic in these compartments.

However, although the current within these compartments can be
optimized, the stationary state of these half-open tubes is always
characterized by a density which approaches either one or zero far
from the open end (with a phase transition
separating these two cases). Long tubes are therefore either
essentially filled or essentially empty. In axons which are
mimicked by this tube geometry, additional regulatory processes are
therefore necessary to maintain efficient stationary transport.

These effects are not restriced to the tube geometry but apply generally
to systems which are coupled to a particle reservoir at one end and
which have two types of channels
'perpendicular' to this surface, one with unsymmetric directed
and one with symmetric diffusive transport.

\appendix
\section{The closed tube}\label{secClosed}

So far, the motion of motors in a half-open geometry has been
considered. A related system is one with closed boundaries at both
ends which has been studied in
\cite{LKN2001,KlumppNieuwenhuizenLipowsky2005}. In this
case, motors cannot enter or leave the system, and the total
number of motors within the tube is fixed:
\begin{equation}\label{particleConservation0}
    N = \int_{0}^L \left[\rhoB(x) + \phi \rhoUb(x)\right]\, dx.
\end{equation}
For this closed system, the same arguments as for the half-open
system can be applied, however the integral constraint
(\ref{particleConservation0}) makes the analysis more complicated.

\begin{figure}
\includegraphics[width=13cm]{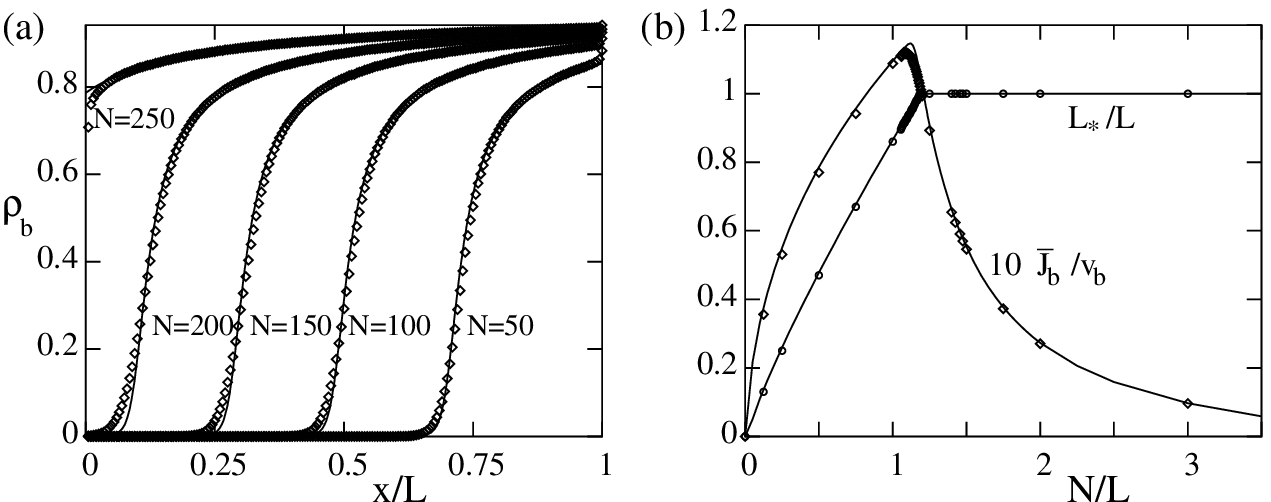}
\caption{%
(a) Density profiles $\rhoB$ for the closed tube for various
particle numbers $N$ as obtained from Monte Carlo simulations
($\opendiamond$) and from the analytical jam
region approximation (\ref{implicitRhoClosed}) (solid lines). The
parameters are $\vb=1/300$, $\eps=2/300$, $\piad=2/3$
and $\phi=4$. \newline
(b) Jam length $L_{*}$ ($\opendiamond$) and average bound current $\JB$
($\opencircle\hspace{-1ex}$) for the closed tube as functions of the number $N$
of motors within the tube. The data points are from simulations, the lines
display the corresponding results from the adsorption equilibrium
approximation as given by equations (\ref{implicitRhoClosed}) and
(\ref{JbtotClosed}), respectively. The other parameters are as
in (a). %
} \label{figClosed}
\end{figure}

Typical profiles for the closed system and motors moving to the
right ($\vb>0$) are displayed in figure \ref{figClosed}(a). The
profiles show the same two characteristic length scales as in the
case of the half-open system: the bulk length $\xi$ and the jam
length $L_*$. The bulk length $\xi$ which is independent of the
boundary densities is again given by equation (\ref{xi}).

Depending on their direction of motion, the motors jam up at one
of the closed ends. In this region, the motors move slowly, and the
adsorption equilibrium approximation (\ref{adsorpEquil2}) can be
applied which leads again to an implicit equation for the bound
density:
\begin{equation}\label{implicitRhoClosed}
      \frac{x}{L} = \frac{1}{V}\left[ g(\rhoB(x),K) - g(\rhoB(0),K) \right]
\end{equation}
with the function $g$ as defined in equation (\ref{gfunc}). This
equation corresponds to equation (\ref{implicitRho}), however, the
left boundary density $\rhoB(0)$ is now unknown. It can be
determined from the particle conservation constraint
(\ref{particleConservation0}), which for adsorption equilibrium
can be integrated and leads to:
\begin{equation}\label{ParticleConservation}
    V\frac{N}{L(1+\phi)} = F(\rhoB(L),K,\phi) - F(\rhoB(0),K,\phi)
\end{equation}
with the function
\begin{eqnarray}
F(\rho,K,\phi) =
\frac{1}{K}\ln\left[\frac{\left|\frac{1}{1-K}-\rho\right|}{1-\rho}\right]
        - \frac{1}{1 - (1-K)\rho}\left[ 1
        + \frac{K}{2}\frac{\phi}{1+\phi}\frac{1}{1 - (1-K)\rho}\right].
\end{eqnarray}
Together with
\begin{equation}\label{ParticleConservation2}
     V = g\left(\rhoB(L),K)\right) - g\left(\rhoB(0),K\right)
\end{equation}
(obtained from equation (\ref{implicitRhoClosed}) by setting
$x=L$) one has two nonlinear equations for the boundary densities
$\rhoB(0)$ and $\rhoB(L)$, which are needed in equation
(\ref{implicitRhoClosed}). The resulting density profiles agree
well with the Monte Carlo profiles in the jam region, see figure
\ref{figClosed}(a).

The average bound current and its maximum can be calculated in the
same way as for the half-open tube, leading to:
\begin{equation}\label{JbtotClosed}
    \frac{\JB}{\vb} = \frac{1}{V} \frac{K}{(1-K)^2}\left[ \frac{1}{\frac{1}{1-K}-\rhoB(L)} - \frac{1}{\frac{1}{1-K}-\rhoB(0)}\right]
\end{equation}
with an extremum as function of the particle number $N$ at
\begin{equation}
    \jB\left(\rhoB(0)\right) = \jB\left(\rhoB(L)\right) \quad \mathrm{or} \quad \rhoB(0) =
    1-\rhoB(L).
\end{equation}
Thus, from equation (\ref{ParticleConservation}), the maximum
occurs for the particle number
\begin{equation}
    \frac{N^{\max}}{L(1+\phi)} = \frac{1}{V} \left[ F(1-\rhoB(L)) - F(\rhoB(L))\right].
\end{equation}
The results for the average current agree quite well with the
simulation results, see figure \ref{figClosed}(b). Only near the
current extremum the Monte Carlo curve is slightly sharper than
the curve from the adsorption equilibrium approximation which
leads to a different value for the current extremizing particle number
$N^{\max}$.

For a very long tube, the current extremum occurs at
particle number
\begin{equation}
\frac{N^{\max}}{L(1+\phi)}\approx \frac{1}{1+K^2} \left( 1+ \frac{1}{L}\frac{\Dub\phi}{\vb}h(K,\phi)\right)
\end{equation}
with $h(K,\phi) = \frac{(1-K)^2}{K} -\frac{1}{K}\ln{K}-\frac{(1-K)(1+K^2)}{2K}\frac{2+(3+K)\phi}{1+\phi} $,
and thus approaches $\frac{1}{1+K^2}$ for large $L$.


\end{document}